     \documentstyle[12pt]{article}   
     \newlength{\dinwidth}                       
     \newlength{\dinmargin}                      
\def\Journal#1#2#3#4{{#1} {\bf #2}, #3 (#4)}

\newcommand{\eabe} {\begin{eqnarray}}
\newcommand{\eaen} {\end{eqnarray}}
\newcommand{\eqbe} {\begin{equation}}
\newcommand{\eqen} {\end{equation}}
\newcommand{\mrm} {\mathrm}
\newcommand{\srm}[1] {_{\mathrm{#1}}}

\renewcommand{\ln} {\mrm {ln}}
\newcommand{\ee} {e^+e^-}

\newcommand{\pt} {p_\perp}

\newlength{\figwidth}
\def\figdir{./figs}

\include{psfig}

\def\lsim{\mathrel{\rlap{\lower4pt\hbox{\hskip1pt$\sim$}}
    \raise1pt\hbox{$<$}}}                
\def\gsim{\mathrel{\rlap{\lower4pt\hbox{\hskip1pt$\sim$}}
    \raise1pt\hbox{$>$}}}                
\begin{document}
\begin{titlepage}
\begin{flushright}
NORDITA 1999/72 HE
\end{flushright}
\vspace*{10mm}
\begin{center}  \begin{Large} \begin{bf}
Comparing Quark Jets in e$^+$e$^-$ and DIS

  \end{bf}  \end{Large}
  \vspace*{5mm}
  \begin{large}
Patrik Ed\'en\\
  \end{large}
NORDITA, Blegdamsvej 17, DK-2100 Copenhagen, Denmark
\end{center}
\vspace*{3cm}
\begin{quotation}
\noindent
{\bf Abstract:} I discuss event cuts in DIS which improve the
similarity between the current Breit hemisphere and hemispheres in
$e^+e^-$ annihilation. I also present a method to study the scale
evolution in quark hemispheres using data from fixed energy $e^+e^-$
experiments. This method can benefit from the high statistics and
flavour tagging at LEP1, also at scales relevant for a comparison with
HERA results.
\end{quotation}
\end{titlepage}

\section{Introduction}\label{sec:intro}
In deep inelastic scattering (DIS), the current hemisphere of the Breit frame is expected to be very similar to one hemisphere in an $e^+e^-$ experiment. This expectation relies on the fundamental assumption of quark fragmentation universality. 
However, a set of features in DIS introduce corrections to this assumption.
In this letter I will give an abbreviated presentation of the investigations in~\cite{breit}, adressing two main issues:

QCD radiation can give rise to high-$p_\perp$ emissions, without correspondance in an $e^+e^-$ event. A high-$p_\perp$ emission in DIS sometimes manifests itself with a completely empty current region~\cite{empty}. A sizeable rate of such events introduces uncertainties to the interpretation of the data. Furthermore, problems arise even if the current region is not empty, but merely depopulated. It is therefore of interest to find a way to exclude high-$p_\perp$ events using another signal than an empty current Breit hemisphere. In sections~\ref{sec:jets} and~\ref{sec:CBHres} we discuss an approach based on jet reconstruction.

The flavour composition in $e^+e^-$ and DIS is not exactly the same. When excluding high-$p_\perp$ events from the DIS analysis, the boson-gluon fusion channel for heavy quark production is supressed.
This implies a lower heavy quark rate in the studied DIS sample, as compared to $e^+e^-$ data at corresponding energies. A uds enriched $e^+e^-$ data sample with high statistics is available from the ${\mrm Z}^0$ peak.
In section~\ref{sec:eeuds} we discuss a method to study properties of $e^+e^-$ quark hemispheres at different scales, using data from the fixed energy ${\mrm Z}^0$ experiments.

\section{Jet Algorithms}\label{sec:jets}
In order to exclude from the DIS sample high-$\pt$ events without correspondance in $\ee$, it is natural to use $k_\perp$-type cluster algorithms with a jet resolution set to $Q/2$. This scale is in analogy to the kinematical constraint for gluon emissions in $e^+e^-$, which is $p_{\perp{\mrm g}}\le E\srm{g}\le\sqrt{s}/2$.

In the HERA experiments, particles which in the lab frame have  a large pseudo-rapidity w.r.t.\ the proton direction are not detected. 
In our analysis, we have chosen to exclude all particles with pseudo-rapidity larger than $3.8$, to take this into account.
For clustering purposes, an initial cluster is introduced along the proton direction, carrying the missing longitudinal momentum.

Historically,  $k_\perp$ algorithms were first designed for $e^+e^-$ physics, and a set of different algorithms exist~\cite{jetalgs}. In DIS, jet observables depend on structure functions and an algorithm where the jet properties factorizes into perturbatively calculable coefficients convoluted with the structure functions is in general preferred. A $k_\perp$ algorithm carried out in the Breit frame and designed to fulfill jet requirements in DIS is presented in~\cite{ktalg}, but we have chosen not to use it in the present investigation for the following reasons:  

We do not intend to investigate jet cross sections or the specific jets found by the cluster algorithm, but merely to exclude high-$p_{\perp}$ events. In the accepted sample, the analysis is performed on the current Breit hemisphere, which is defined independently of any jet reconstruction scheme. Thus the infrared properties of the used algorithm, and in particular the problems adressed by the Breit frame algorithm in~\cite{ktalg}, are not essential for the present study.

It is in~\cite{ktalg} stressed that the appealing benefits of the algorithm requires the jet resolution scale $E_t$ to satisfy $\Lambda_{QCD}^2\ll E_t^2 \ll Q^2$. By setting $E_t=Q/2$, we would in the present study use the Breit frame algorithm in a way for which it is not intended.

For these reasons, we have chosen to use the simpler $e^+e^-$ $k_\perp$ algorithms.
To estimate the reliability of the results obtained using jets, we have used three different algorithms, \textsc{Luclus}~\cite{jetset}, \textsc{Durham}~\cite{durham} and \textsc{Diclus}~\cite{diclus}, applied in the hadronic CMS.

\begin{figure}[tb]
\parbox{0.47\textwidth}{
	\mbox{\psfig{figure=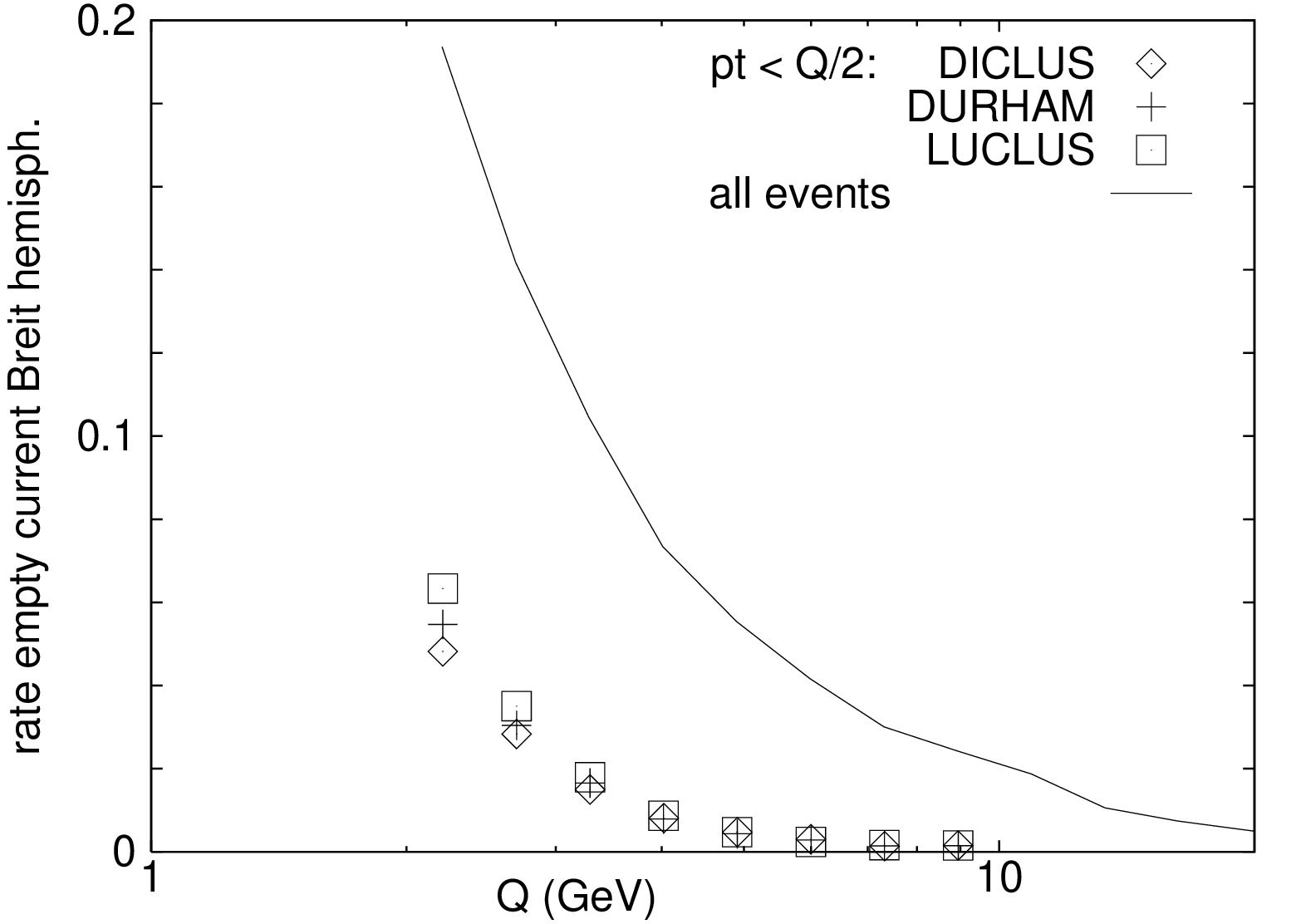,width=0.46\textwidth}}
\caption{\em The fraction of MC simulated events with empty current Breit hemispheres is significantly reduced after a cut in jet-$p_\perp$.}
  \label{f:empty}
}
~
\parbox{0.47\textwidth}{
	\psfig{figure=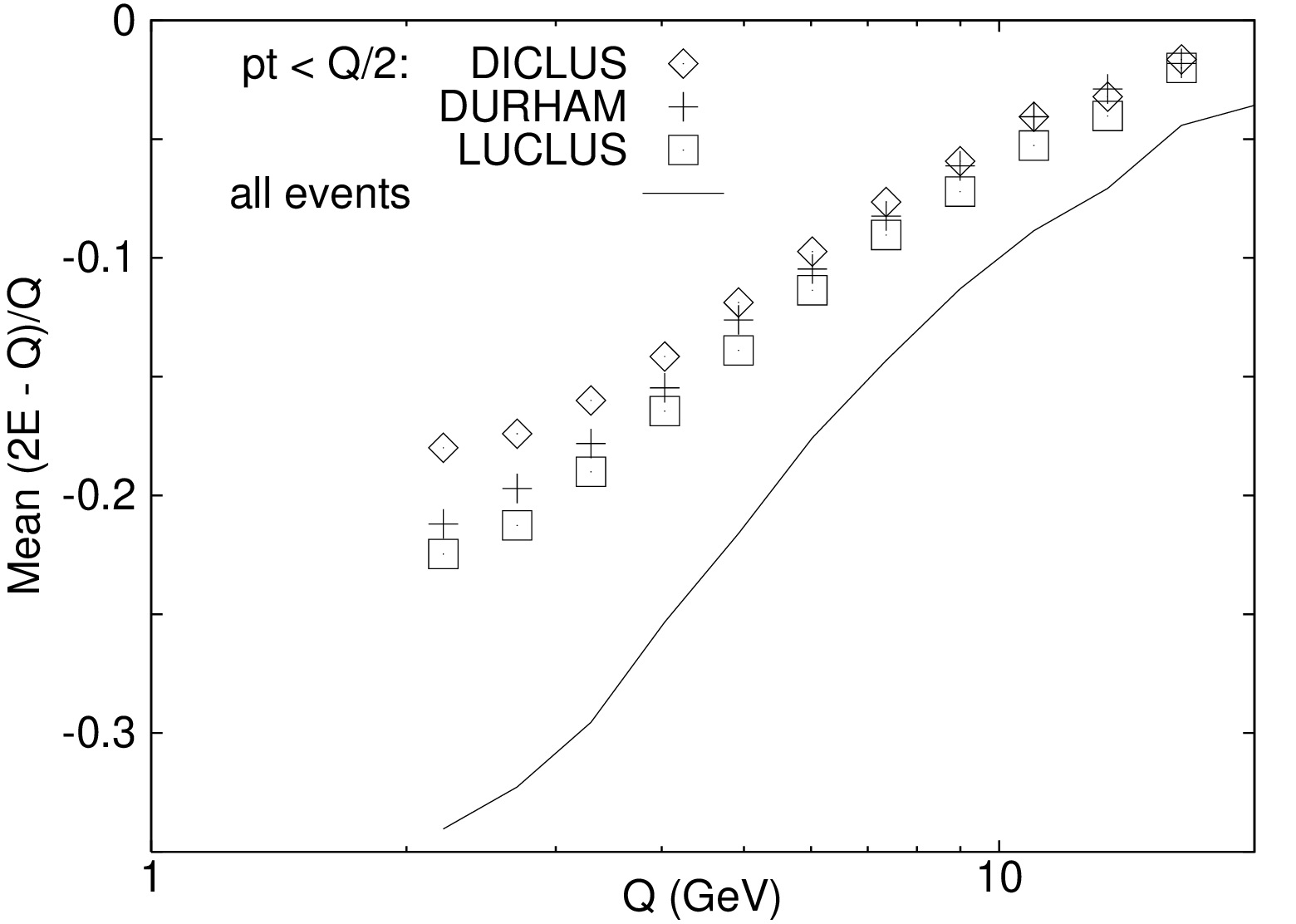,width=0.46\textwidth}
\caption{\em The relative difference between mean energy in the current Breit hemisphere and $Q/2$ is reduced after a  cut in jet-$p_\perp$.}
  \label{f:Ediff}
}
\end{figure}

\section{Current Breit Hemisphere Properties}\label{sec:CBHres}
To examine if a cut in jet-$p_\perp$ is suitable to isolate a sample where universality is expected to hold, we study results for the current Breit hemisphere in MC generated events at HERA energies. The results are compared to $e^+e^-$ MC results.

We simulate the electro-weak interaction in DIS  using the \textsc{Lepto} MC~\cite{lepto}. In both the $e^+e^-$ and DIS simulations, we use the Colour Dipole Model~\cite{CDM}, implemented in \textsc{Ariadne}~\cite{ariadne}, to describe the parton cascade. The  Lund string fragmentation model~\cite{stringmodel}, implemented in \textsc{Jetset}~\cite{jetset} is then used to describe hadronization. The chosen MC programs give a generally good description of data in DIS and $e^+e^-$.

The kinematical region of the DIS simulation is constrained by the cut $Q^2>4\mrm{GeV}^2$. For each $Q^2$, the whole range of possible $x$-values is considered. Ideally, the properties of the current Breit hemisphere would depend on the scale $Q^2$ only, but QCD radiative corrections introduces an implicit $x$ dependence.
When high-$p_\perp$ events are excluded, the implicit $x$ dependence is strongly reduced. To show that radiative corrections to the properties of the current Breit hemisphere are reduced by the cuts, it will suffice to consider the results integrated over $x$.

\subsubsection*{Effects of the $p_\perp$-cut}
In Fig~\ref{f:empty}, the fraction of events in DIS with an empty current Breit hemisphere is shown to be significantly reduced after a $p_\perp$$<Q/2$ cut, for all jet finding algorithms. 

Another interesting observable is the average energy in the current Breit hemisphere, which in general differs from a corresonding $e^+e^-$ hemisphere. Consider two hemispheres in the rest frame of an $e^+e^-$ annihilation event. In general, they have different masses and hence different energies. In other words, a high-$p_\perp$ emission in one hemisphere reduces the energy of the other, due to energy--momentum conservation.

As for an $e^+e^-$ hemisphere, the energy of the current Breit hemisphere at fixed $Q^2$ depends on emissions in this hemisphere and in the nearby phase space of the opposite hemisphere. The latter is very large in DIS at low $x$, but after our suggested event cuts, it is reduced by the condition $p_\perp$$<Q/2$. The corresponding kinematical constraint in $e^+e^-$ is however $\left| \mathbf{p}\right|<Q/2$, which is more restrictive. Thus the region for high-$p_\perp$ emissions which can reduce the energy of the considered hemisphere is larger in DIS than in $e^+e^-$. Without reaching for a quantitative prediction, we expect the mean energy and multiplicity of the current Breit hemisphere to satisfy
\eqbe \left<E\srm{CBH}\right> < \frac Q 2,~~~\left<N\srm{CBH}\right> < \frac1 2N_{ee}(Q^2), \eqen
where $N_{ee}(Q^2)$ is the average multiplicity in an $e^+e^-$ experiment with invariant mass $Q^2$.

If $2\left<E\srm{CBH}\right>/Q$ would be much smaller than 1, the similarity between the current Breit hemisphere and an unbiased $e^+e^-$ hemisphere is poor, and it is not likely that reliable conclusions from a comparison can be drawn. On the other hand, if $2\left<E\srm{CBH}\right>/Q \approx 1$, the current Breit hemisphere sample may be closely related to an unbiased $e^+e^-$ hemisphere.
In Fig~\ref{f:Ediff}, the relative energy shift $2\left<E\srm{CBH}\right>/Q - 1$ is shown to be sizeable for the unrestricted event sample, but significantly reduced after imposing a $p_\perp < Q/2$ cut.
 
\subsubsection*{Flavour Compositions}
A cut in $p_\perp$ suppresses the boson-gluon fusion channel for charm production in DIS and reduces the charm rate. In our MC simulations, charm is suppressed  by roughly a factor 2 in a $p_\perp< Q/2$ event sample.
A problem for the interpretation of this result is the fact that the MC simulation program significantly underestimates the rate of charm events. In the MC this is about 10\%, while data indicate a much larger value around 25\%~\cite{charmrates}.
However, a similar suppression ought to be expected also for real data. If  boson-gluon fusion is a more important source of charm in nature than assumed in the MC model, the suppression may be even stronger.  It would then be appropriate to compare with uds enriched $e^+e^-$ data. How to obtain such data at different energies will be discussed in section~\ref{sec:eeuds}.

\begin{figure}[tb]
  \begin{center}  \hbox{ \vbox{
	\mbox{\psfig{figure=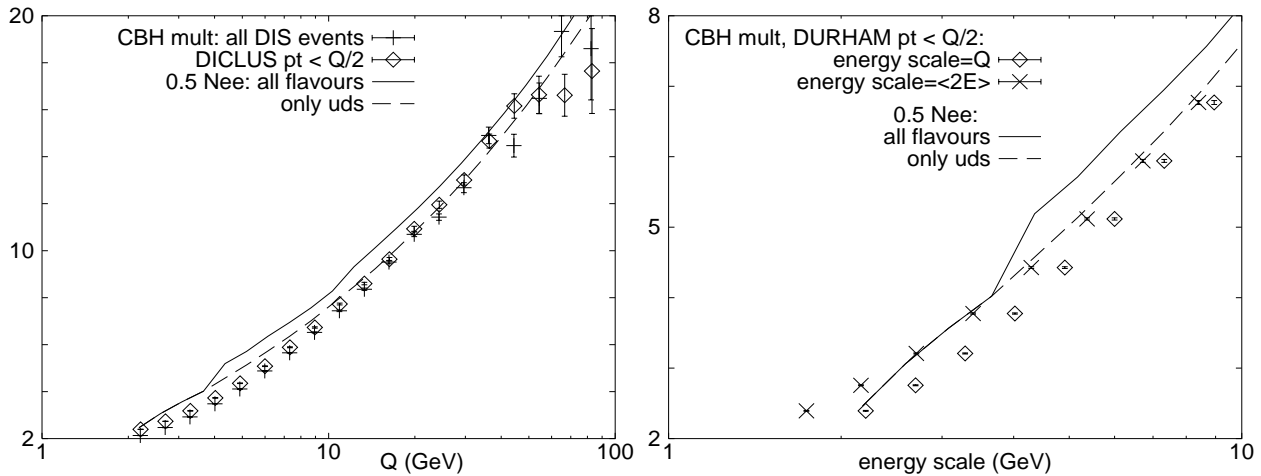,width=0.98\textwidth}}
   }  }
  \end{center}  \caption{\em MC simulations of multiplicities in $e^+e^-$ and current Breit hemispheres (CBH). {\bf LEFT:} $e^+e^-$ results with $s=Q^2$, allowing for all flavours (solid line) and only uds (dashed line), compared to  CBH multiplicities using all events (crosses) and a jet-$p_\perp < Q/2$ sample (diamonds). Both the exclusion of high-$p_{\perp}$ DIS events and heavy flavour $e^+e^-$ events improves the agreement between the multiplicities. {\bf RIGHT:} For low scales, the agreement is further improved if comparing the CBH multiplicity with $e^+e^-$ data with squared mass $s=4\left<E\srm{CBH}\right>^2$.}
  \label{f:multuds}
\end{figure} 
Monte Carlo results for average multiplicities are presented in the left plot of Fig~\ref{f:multuds}. The results for the  low-$p_\perp$ DIS sample are essentially equivalent for the three considered jet algorithms, and only the \textsc{Diclus} result is shown. 
In this MC generated event sample, heavy quark rates are well below 10\%, and it is therefore better compared to a MC $e^+e^-$ sample with only uds. 
The agreement between the low-$p_{\perp}$ and the $e^+e^-$ uds sample is better than between the unrestricted event samples also shown in Fig~\ref{f:multuds}. 

\subsubsection*{Energy Scale Corrections}
As discussed previously in this section, high-$p_\perp$ emissions close in rapidity to the considered hemisphere implies a reduction of the average energy in the current Breit hemisphere, $\left<E\srm{CBH}\right>$, to a value slightly smaller than $Q/2$. It is then natural to compare the current Breit hemisphere multiplicities with $e^+e^-$ data at a squared mass $s=4\left<E\srm{CBH}\right>^2$. The agreement between $e^+e^-$ and Breit frame multiplicities is then significantly improved, as shown in the right plot of Fig~\ref{f:multuds}.  Only the \textsc{Durham} result is presented, but the other results are very similar.

Though the focus here is on radiative corrections, we can not firmly exclude the possibility that there are differences between $\ee$ and DIS also in the hadronization phase, which could imply an non-perturbative flow of energy and particles between the current region and the target region. Such an effect is masked by the radiative corrections, and  an unexpected breaking of quark fragmentation universality in the hadronization phase could hardly be seen in the average multiplicity. However, once expected differences are corrected for, unexpected features which break quark fragmentation universality can still be searched for in less blunt observables, like strangeness rates~\cite{s_anomaly}, energy spectra and higher multiplicity moments.

Comparing the left and right plot in Fig~\ref{f:multuds}, we see that the high-$p_\perp$ cut has a relatively small influence on the average multiplicity, as compared to the energy scale shift. However, in Fig~\ref{f:empty} we note that the rate of empty current Breit hemisphere events, which are excluded from the analysis more by necessity than by theoretical understanding, are reduced with the high-$p_\perp$ cut. This is also the case for the energy shift, as seen in Fig~\ref{f:Ediff}. Thus we find that a cut in jet-$p_\perp$ is a powerful step towards an event sample where quark fragmentation universality is expected to hold.

\section{Scale Evolutions in Fixed Energy $\mathbf{e^+e^-}$ Annihilation}\label{sec:eeuds}
A large sample of uds enriched events are available from LEP1 at $\sqrt s=90$GeV. This is not the case for other energies corresponding to the HERA kinematical range.
In this section we discuss how the scale evolution of $e^+e^-$ uds hemispheres can be examined, using data from a fixed energy $e^+e^-$ experiment.

In an $e^+e^-$ experiment at squared mass $s$, we consider an artificial scale $Q\srm{max}^2<s$. Three jets are reconstructed with a $k_\perp$ cluster algorithm, and events where $p_\perp$$> Q\srm{max}/2$ are excluded. 
We study the multiplicity for particles where the rapidity, measured in the thrust direction, satisfies
\eqbe \left|y\right|>\frac1 2 \ln(s/Q\srm{max}^2). \label{e:tcone} \eqen  
These cones in the thrust direction correspond to $e^+e^-$ hemispheres with squared invariant mass $Q\srm{max}^2$, and their evolution with this scale can thus be studied in a fixed energy experiment. (For a more detailed discussion, see e.g.~\cite{jetscales}.)

This ``thrust-cone'' method is designed to be similar to the Breit frame analysis. Two differences are present, due to the absence of a $t$-channel probe in $e^+e^-$. The scale $Q^2$ and the direction which defines the quark jets are determined by the probe in DIS. In  $e^+e^-$, the scale $Q\srm{max}^2$ is chosen freely and the rapidity is measured along the thrust direction. It is possible to use the thrust direction also in the DIS analysis, which could improve the similarity with the $\ee$ thrust-cone algorithm. A thorough investigation of the quantitative effects of this adjustment is however better performed with a full detector simulation, as they may depend on how the thrust direction is reconstructed in DIS events with a large fraction of the target region undetected.

\begin{figure}[tb]
  \begin{center}  \hbox{
     \vbox{
	\mbox{ \psfig{figure=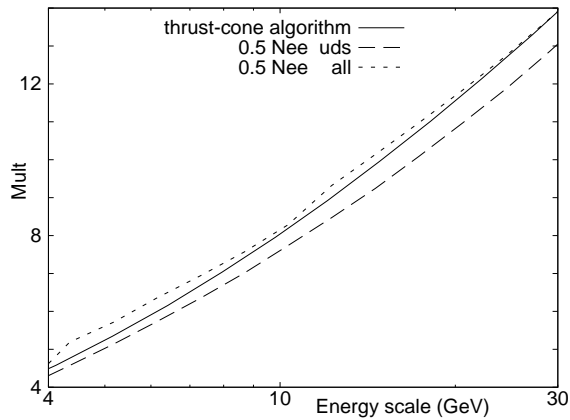,width=\figwidth}	}
    }
  }
  \end{center}  \caption{\em Multiplicities in MC simulated $e^+e^-$ events including all allowed flavours (dotted line) are higher than for light quarks events (dashed line). The multiplicity obtained by the thrust-cone algorithm in a uds sample at the ${\mrm Z}^0$ pole (solid line) are in better agreement with the uds result for moderate scales between 4 and 8 GeV. This corresponds to current Breit hemispheres in DIS with $Q^2$ between 16 and 64 $\mrm{GeV}^2$. For larger scales the thrust-cone results differ from the uds sample in a similar way as the full flavour results.}
  \label{f:eescales}
\end{figure}
In Fig~\ref{f:eescales} the multiplicity evolution of an $e^+e^-$ uds sample is compared to results including all flavours and results obtained by the thrust-cone algorithm in a fixed energy uds sample. The thrust-cone multiplicities are plotted as a function of the average energy scales obtained in the cones. These scales are somewhat smaller than the chosen $Q\srm{max}/2$, and the deviations are of order  5 to 10\%.

The aim of the thrust-cone algorithm is to reproduce the scale evolution of a $e^+e^-$ uds sample  better than a sample including all flavours. As seen in Fig~\ref{f:eescales}, this is achieved for moderate $Q\srm{max}$, between 4 and 8 GeV. This corresponds to current Breit hemispheres in DIS with $Q^2$ between 16 and 64 $\mrm{GeV}^2$, which is an important range at the HERA experiments. 

In rare cases where an emitted gluon gives the hardest jet, the thrust will be directed along the gluon momentum. Thus the average multiplicity in the thrust-cone analysis will get some contribution from gluon jets, which results in a systematic overestimation of the multiplicity. 
For $Q\srm{max} >$ 8 GeV, the effect of gluon ``pollution'' is of the same order as the effect of the heavy quarks. 
However, in a realistic experimental analysis, the thrust-cone results applied on uds events at the ${\mrm Z}^0$ peak will benefit from very large statistics compared to experiments at lower energies, and also from the fact that the scale evolution can be studied over a large range with the same detector. To conclude this section, our investigation indicates that it would be interesting to compare DIS data not only to full $e^+e^-$ results at different $s$, but also with thrust-cone results in uds samples from the ${\mrm Z}^0$ experiments.

\section{Summary}\label{sec:summary}
The assumption of quark fragmentation universality implies that the current Breit hemisphere in DIS is expected to be very similar to a hemisphere in $e^+e^-$ annihilation. However, the experimental situations are different, and several corrections to universality are present. We have here proposed event cuts to improve the expected validity of quark fragmentation universality. 

We suggest to exclude from the DIS analysis events with high-$p_\perp$ emissions, which have no correspondence in $e^+e^-$ events.  The $p_\perp$-scale can be reconstructed using jet cluster algorithms with the resolution scale $Q/2$. For these purposes, it is suitable to use $k_\perp$ clustering algorithms originally designed for analyses of $\ee$ annihilation events.
Using Monte Carlo simulations, we investigate three different types of $k_\perp$ cluster schemes, and find that the agreement between $e^+e^-$ and Breit frame results improves after a cut in jet $p_\perp$, independently of the specific choice of $\ee$ $k_\perp$ algorithm.

In the accepted low-$p_\perp$ DIS sample, heavy quarks are suppressed. This motivates a comparison with uds enriched $e^+e^-$ data, which are available from the experiments at the ${\mrm Z}^0$ pole, but not at lower energies. We have here presented a method, the ``thrust-cone algorithm'', to study scale evolutions of $e^+e^-$ quark hemispheres, using data from fixed energy experiments. With this algorithm, uds enriched data with high statistics from the LEP1 experiments can be compared to results for the current Breit hemisphere, over a large range of energies.

\subsection*{Acknowledgments}
I thank Leif L\"onnblad and G\"osta Gustafson for their  significant contributions to this investigation.

\end{document}